\input harvmac

\def\ls{l_s}
\def\ms{m_s}
\def\gs{g_s}
\def\lpx{l_P^{10}}
\def\lpii{l_P^{11}}
\def\R11{R_{11}}
\def\mb{{m_{\rm brane}}}
\def\vb{{v_{\rm brane}}}

\def\half{{1 \over 2}}
\def\identity{{\rlap{1} \hskip 1.4pt 1}}
\def\laplace{{\kern1pt\vbox{\hrule height 1.2pt\hbox{\vrule width 1.2pt\hskip
  3pt\vbox{\vskip 6pt}\hskip 3pt\vrule width 0.6pt}\hrule height 0.6pt}
  \kern1pt}}
\def\scriptlap{{\kern1pt\vbox{\hrule height 0.8pt\hbox{\vrule width 0.8pt
  \hskip2pt\vbox{\vskip 4pt}\hskip 2pt\vrule width 0.4pt}\hrule height 0.4pt}
  \kern1pt}}

\newcount \pageit
\footline={\tenrm\hss \ifnum\pageit=0 \hfill \else \number\pageno \fi\hss}
\pageit=0
\pageno=0
\font\titlerm=cmr10 scaled \magstep3

\line{\hfill hep-th/9603127}
\line{\hfill RU--96--17}
\line{\hfill March 1996}
\bigskip
\bigskip
\centerline{\titlerm A Comment on Zero-brane Quantum Mechanics}
\vskip 24pt
\centerline{Daniel Kabat and Philippe Pouliot}
\vskip 8pt
\centerline{\it Department of Physics and Astronomy}
\centerline{\it Rutgers University}
\centerline{\it Piscataway, NJ 08855--0849}
\vskip 8pt
\centerline{\tt kabat, pouliot@physics.rutgers.edu}
\vskip 0.9 true in
\leftskip = 0.5 true in
\rightskip = 0.5 true in
\noindent
We consider low energy, non-relativistic scattering of two Dirichlet
zero-branes as an exercise in quantum mechanics.  For weak string
coupling and sufficiently small velocity, the dynamics is governed by
an effective U(2) gauge theory in 0+1 dimensions.  At low energies,
D-brane scattering can reliably probe distances much shorter than the
string scale.  The only length scale in the quantum mechanics problem
is the eleven dimensional Planck length.  This provides evidence for
the role of scales shorter than the string length in the weakly
coupled dynamics of type IIA strings.
\leftskip = 0.0 true in
\rightskip = 0.0 true in
\vfill
\eject
\pageit=1

\def\np{{\it Nucl.~Phys.~}}
\def\pl{{\it Phys.~Lett.~}}

\def\mpl{{\it Mod.~Phys.~Lett.~}}

\nref\DaiLeiPol{J.~Dai, R.~G.~Leigh, and J.~Polchinski, \mpl
{\bf A4} (1989) 2073.}
\nref\Leigh{R.~G.~Leigh, \mpl {\bf A4} (1989) 2767.}
\nref\PolBdy{J.~Polchinski, {\it Combinatorics of Boundaries in String
Theory}, hep-th/9407031.}
\nref\Polchinski{J.~Polchinski, {\it Dirichlet-Branes and Ramond-Ramond
Charges}, hep-th/9510017.}
\nref\WittenI{E.~Witten, {\it Bound States of Strings and p-Branes},
hep-th/9510135.}
\nref\Li{M.~Li, {\it Boundary States of D-Branes and Dy-Strings},
hep-th/9510161\semi
{\it Dirichlet Boundary State in Linear Dilaton Background},
hep-th/9512042.}
\nref\KleTho{I.~R.~Klebanov and L.~Thorlacius, {\it The Size of p-Branes},
hep-th/9510200.}
\nref\Bachas{C.~Bachas, {\it D-Brane Dynamics}, hep-th/9511043.}
\nref\CalKle{C.~G.~Callan, Jr.~and I.~R.~Klebanov,
{\it D-Brane Boundary State Dynamics}, hep-th/9511173.}
\nref\BanSus{T.~Banks and L.~Susskind, {\it Brane-Antibrane Forces},
hep-th/9511194.}
\nref\BerVafSad{M.~Bershadsky, C.~Vafa, and V.~Sadov, {\it D-Branes
and Topological Field Theories}, hep-th/9511222.}
\nref\Town{P.~K.~Townsend, {\it D-branes from M-branes}, hep-th/9512062.}
\nref\Douglas{M.~R.~Douglas, {\it Branes within Branes},
hep-th/9512077.}
\nref\KhuMye{R.~R.~Khuri and R.~C.~Myers, {\it Low-Energy Scattering
of Black Holes and $p$-branes in String Theory}, hep-th/9512137.}
\nref\Schmi{C.~Schmidhuber, {\it D-brane actions}, hep-th/9601003.}
\nref\Gub{S.~S.~Gubser, A.~Hashimoto, I.~R.~Klebanov, and J.~M.~Maldacena,
{\it Gravitational lensing by $p$-branes}, hep-th/9601057.}
\nref\PolChaJoh{J.~Polchinski, S.~Chaudhuri, C.~V.~Johnson,
{\it Notes on D-Branes}, hep-th/9602052.}
\nref\TownII{P.~K.~Townsend, \pl {\bf B350} (1995) 184, hep-th/9501068.}
\nref\WittenII{E.~Witten, {\it String Theory Dynamics in Various
Dimensions}, hep-th/9503124.}
\nref\Shenker{S.~H.~Shenker, {\it Another Length Scale in String Theory?},
hep-th/9509132.}
\nref\FraTse{E.~S.~Fradkin and A.~A.~Tseytlin, \pl {\bf B163} (1985)
123.}
\nref\Abou{A.~Abouelsaood, C.~G.~Callan, C.~R.~Nappi, and S.~A.~Yost,
\np {\bf B280} (1987) 599.}
\nref\Tse{A.~A.~Tseytlin, {\it Self-Duality of Born-Infeld action and
Dirichlet 3-brane of type IIB superstring theory}, hep-th/9602064.}
\nref\GSW{M.~B.~Green, J.~H.~Schwarz, and E.~Witten, {\it Superstring
Theory} (Cambridge University Press, 1987) p.~288.}
\nref\DanFerSun{U.~H.~Danielsson, G.~Ferretti, and B.~Sundborg, {\it
D-particle Dynamics and Bound States}, hep-th/9603081.}

Dirichlet-branes \refs{\DaiLeiPol{--}\PolBdy} in string theory have
been intensively studied since the realization that they carry
Ramond-Ramond charge and are responsible for many of the
non-perturbative effects required by duality
\refs{\Polchinski{--}\Gub}; for a recent review see \PolChaJoh.  At
low energies -- energies below the string scale -- the dynamics of
D-branes should be governed by the effective theory of the massless
modes of the open strings which end on the branes.  For a single
$p$-brane, this effective theory is the reduction of a ten dimensional
N=1 supersymmetric abelian gauge theory to the $p+1$ dimensional
world-volume of the brane.  The $p+1$ components of the gauge field
tangent to the brane give rise to an internal gauge theory on the
world-volume, while the $9-p$ scalars which arise from components of
the gauge field normal to the brane are interpreted as position
coordinates that translate the brane \refs{\DaiLeiPol,\Leigh}.  When
$n$ branes are present, open strings carry a Chan-Paton factor to
label which brane they end on, and the effective gauge theory gets
enlarged to a $U(n)$ gauge group.  This description has been used to
analyze bound states of D-branes \WittenI.

We consider Dirichlet zero-branes, or D-particles, in ten dimensional
type IIA string theory.  The IIA string can be obtained by
compactifying M-theory on a circle \refs{\TownII, \WittenII}, and the
low energy effective theory of strongly coupled type IIA string theory
is known to be eleven dimensional supergravity.  The zero-branes we
study have an interpretation as Kaluza-Klein modes of the eleven
dimensional theory.  This suggests that eleven dimensional physics
should play a role in their dynamics; we will indeed find evidence for
this below.

One of the more intriguing hints from duality is the appearance of new
length scales in string theory \Shenker.  In ten dimensions, the IIA
string is characterized by the string length $\ls = 1/\ms$ and the
string coupling $\gs$.  A dual M-theory description is
characterized, at least at low energies, by the eleven dimensional
Planck length $\lpii$ and the compactification radius $\R11$.  By
matching the low energy supergravity theories one finds $\R11 / \lpii
\sim \gs^{2/3}$ \refs{\WittenII}, while by matching the IIA string
tension to the tension of a membrane wrapped around the eleventh
dimension one finds $\ls^2 = (\lpii)^3/\R11$.  This fixes the
hierarchy of scales
$$\eqalign{
&\ls\cr
\lpx &\sim \gs^{1/4} \ls\cr
\lpii &\sim \gs^{1/3} \ls\cr
\R11 &\sim \gs \ls \cr}$$
where we have also listed the ten dimensional Planck length $\lpx$.

We now show that zero-brane dynamics, in a regime
accurately described by a low energy effective theory, can reliably probe
distances much shorter than the string scale.  The mass of a D-brane
is determined by a BPS formula, $\mb \sim \ms/\gs$.  For zero-branes
this matches their interpretation as Kaluza-Klein modes with mass
$\sim 1/\R11$.  Consider scattering two zero-branes at a
characteristic velocity $\vb \sim \gs^\alpha$.  We work at weak
coupling, so that string loops do not affect the dynamics, and choose
an exponent $\alpha > 0$, so that at weak coupling the branes are
moving very slowly.  The kinetic energy of the branes is then
$$
E_{\rm brane} = {1 \over 2} \mb v_{\rm brane}^2 \sim \gs^{2\alpha-1} \ms\,.
$$
For $\alpha > 1/2$ the kinetic energy is extremely small, much smaller
than the string mass, and we are justified in analyzing the dynamics
in a low energy effective theory.  In particular, effects due to
massive string states ($\alpha'$ corrections to the effective theory)
can be neglected.  Although the energy is small, the momentum does not
have to be small.  The momentum
$$
p_{\rm brane} = \mb \vb \sim \gs^{\alpha - 1} \ms
$$
can be much larger than the string scale provided $\alpha < 1$.  Thus,
there is a range of exponents, $1/2 < \alpha < 1$, for which we can
use low energy field theory to accurately describe the dynamics of the
branes, and in which we can probe distances $\sim 1/p_{\rm brane}$
all the way down to $\gs^{1/2} \ls$.

We now formulate the quantum mechanics problem which governs the
dynamics of two zero-branes.  We take the two zero-branes to have
equal Ramond-Ramond $U(1)$ charges, so that when at rest they form a
BPS saturated state.\footnote{*}{Branes with unequal charges will have
radically different dynamics \BanSus.}  The relevant degrees of
freedom are the massless modes of the open strings which are attached
to the brane.  These correspond to the dimensional reduction of an N=1
supersymmetric $U(2)$ gauge theory from 9+1 to 0+1 dimensions.
Namely, on the world-line of the brane we have fields
$$\eqalign{
A_0 &= {i \over 2} \left(A_0^0 \identity + A_0^a \sigma^a\right)\cr
\phi_i &= {i \over 2} \left(\phi_i^0 \identity +
       \phi_i^a \sigma^a\right)\cr
\psi_A &= {i \over 2} \left(\psi_A^0 \identity +
       \psi_A^a \sigma^a\right)\cr}
$$
where $A_0$ is a single-component $U(2)$ gauge field, $i=1,\ldots,9$
labels the adjoint Higgs fields $\phi_i$, $A=1,\ldots,16$ labels
the real adjoint fermions $\psi_A$, and $a = 1,2,3$ is an
$SU(2)$ index.

A generic expectation value for the Higgs fields $\phi_i^a$ will break
the $U(2)$ gauge symmetry down to $U(1) \times U(1)$ at low
energy.  In terms of zero-branes, this corresponds to the fact that
two widely separated zero-branes are described by a $U(1) \times U(1)$
gauge theory, one $U(1)$ factor associated with each brane.  The two
$U(1)$ Higgs fields in the low energy theory are position coordinates
for the branes\footnote{*}{The adjoint representation of $U(1)$ is
trivial, so these Higgs fields are gauge invariant.  Although one
normally expects a gauge field in $d$ dimensions to have $d-2$ degrees
of freedom, upon reduction to 0+1 dimensions an abelian gauge field
has $d-1$ degrees of freedom.  This makes a particle interpretation
possible.}  \refs{\DaiLeiPol,\Leigh}. This follows from the fact that
in the $\sigma$-model describing an open string attached to the brane
$$
{1 \over 4 \pi \alpha'} \int_{\Sigma} \, g_{\mu\nu} \partial_a X^\mu
\partial_b X^\nu h^{ab} + \oint_{\partial \Sigma} \, \phi_i
\partial_\perp X^i + \cdots
$$
the background field $2 \pi \alpha' \phi_i$ couples to the momentum
conjugate to $X^i$, and therefore translates the endpoints of the open
string.

The effective action for these fields arises at leading order from an
open string disk diagram \refs{\Leigh,\FraTse{--}\Tse}.  At low
energies, corresponding to weak field strengths, we expect the
dynamics to be governed by the standard super Yang-Mills action.  In
units where $2\pi\alpha' = 1$ the action is
$$
S = \int dt \, {1 \over 2 \gs} {\rm Tr} \, F_{\mu \nu} F^{\mu \nu}
            - i {\rm Tr} \, \bar\psi \Gamma^\mu D_\mu \psi\,.
$$
The $1/\gs$ reflects the disk origin of this action; we have absorbed
a similar factor in front of the fermion term by rescaling $\psi$.
We have
$$\eqalign{
F_{0i} &= \partial_0 \phi_i + [A_0,\phi_i]\cr
F_{ij} &= [\phi_i,\phi_j]\cr
\noalign{\smallskip}
D_0 \psi &= \partial_0 \psi + [A_0,\psi]\cr
D_i \psi &= [\phi_i,\psi]\cr}
$$
and adopt a basis in which
$$\eqalign{
\Gamma^0 &= \sigma^2 \otimes \identity_{16 \times 16}\cr
\Gamma^i &= i \sigma^1 \otimes \gamma^i\cr}
$$
where $\gamma^i$ are real symmetric $16 \times 16$ matrices given
explicitly in \GSW.

When expanded, the action is a sum of three terms.  The first term
contains the $U(1)$ gauge degrees of freedom.
$$
S_{\rm cm} = \int dt \, {1 \over 2 \gs} \dot\phi_i^0 \dot\phi_i^0
+ {i \over 2} \psi^0 \dot\psi^0
$$
This is an unconstrained system which describes free motion of the
center of mass.  The second term, involving $SU(2)$ degrees of
freedom, governs the relative motion of the two branes.
$$
S_{\rm relative} = \int dt \, {1 \over 2 \gs} \dot\phi_i^a \dot\phi_i^a
+ {i \over 2} \psi^a \dot \psi^a - {1 \over 4 \gs} \vert\phi_i\times\phi_j
\vert ^2 + {i \over 2} \epsilon_{abc} \phi_i^a \psi^b \gamma^i \psi^c
$$
In the final term, $A_0$ appears as a Lagrange multiplier:
$$
S_{\rm constraint} = \int dt \, {1 \over \gs} \epsilon_{abc} A_0^a
\dot\phi_i^b \phi_i^c + {1 \over 2 \gs} \vert A_0 \times \phi_i \vert^2
+ {i \over 2} \epsilon_{abc} A_0^a \psi^b \psi^c\,.
$$
The equations of motion for $A_0$ require $SU(2)$ invariance, which we
impose as a constraint on physical states.

Quantization is straightforward.  In terms of eight complex fermions
$\chi_M^a = {1 \over \sqrt{2}}(\psi_M^a + i \psi_{M+8}^a)$, the
Hamiltonian is
$$\eqalign{
H &= {\gs \over 2} \vert \pi_i \vert^2 + {1 \over 4 \gs} \vert \phi_i
\times \phi_j \vert^2 - \sum_{i = 1}^7 \epsilon_{abc} \phi_i^a \bar \chi^b
\tilde\gamma^i \chi^c \cr
&\quad - \half \epsilon_{abc} \phi_8^a(\chi^b \chi^c - \bar \chi^b
\bar\chi^c) - {i \over 2} \epsilon_{abc} \phi_9^a (\chi^b \chi^c
+ \bar \chi^b \bar \chi^c)\cr}
$$
where $\tilde \gamma^i$ are a set of seven real, antisymmetric $8 \times 8$
matrices given in \GSW, and
$$ 
i[\pi_i^a,\phi_j^b] = \delta_{ij} \delta^{ab} \qquad \qquad
\lbrace\chi_M^a,\bar\chi_N^b\rbrace = \delta_{MN} \delta^{ab} \,.
$$

Without any dynamical analysis, a simple argument shows that the only
length scale intrinsic to this problem is the eleven dimensional
Planck length.  Recall that $\phi$ and $\pi$ are the distance and
momentum measured in string units.  Introduce rescaled fields
$$
\phi = \gs^{1/3} \phi_{11} \qquad \pi = \gs^{-1/3} \pi_{11} 
$$
which are the corresponding quantities measured in eleven dimensional
Planck units.  In terms of these fields, the Hamiltonian (still
measured in string units) has an overall factor of $\gs^{1/3}$, but no
other coupling constant dependence.  This shows that the only length
scale in the problem is the eleven dimensional Planck length, and that
this length scale can be probed with an energy $\sim \gs^{1/3} \ms$
that is well below the string scale.  We therefore expect zero-brane
scattering to exhibit some interesting features at a momentum of order
the eleven dimensional Planck scale.

To gain additional insight, we have analyzed a toy problem: $s$-wave
scattering in the $0+1$ dimensional theory obtained by dimensional
reduction of $2+1$ dimensional $N=1$ $U(2)$ Yang-Mills theory.  The
toy problem has two adjoint Higgs fields and a pair of real adjoint
fermions.  The Hamiltonian for relative motion is
$$
H = {\gs \over 2} \left( \vert \pi_1 \vert^2 + \vert \pi_2 \vert^2
\right) + {1 \over 2 \gs} \vert \phi_1 \times \phi_2 \vert^2
- {i \over 2} \epsilon_{abc} \left(\phi_1^a + i \phi_2^a\right)
\chi^b \chi^c - {i \over 2} \epsilon_{abc} \left(\phi_1^a - i \phi_2^a
\right) \bar\chi^b \bar\chi^c
$$
with the constraint that physical states must be $SU(2)$ invariant.
For a given angular momentum, the bosonic fields can be conveniently
characterized by the two invariants
$$\eqalign{
r^2 &= \vert \phi_1 \vert^2 + \vert \phi_2 \vert^2\cr
\noalign{\smallskip}
\sin 2\theta &= {2 \vert \phi_1 \times \phi_2 \vert \over \vert \phi_1
\vert^2 + \vert \phi_2 \vert^2}\,.\cr}
$$
The angle $\theta$ ranges from $0$ to $\pi/4$.  When $\theta=0$ the
two $SU(2)$ vectors are aligned, in which case $r$ can be interpreted
as the distance between the branes.

We diagonalize the total angular momentum $j=l+s$, where $l$ is the
orbital angular momentum and $s$ is the spin of the branes.  The
Hamiltonian for a given $j$ acts on a four dimensional space of
fermion states.  Written in a convenient basis, the Hamiltonian for
$s$-wave ($j=0$) scattering is
$$
H = \left(\matrix{
& h_{l=0} & -{i\over\sqrt{2}}re^{i\theta} 
& {i\over\sqrt{2}}re^{-i\theta} & 0 \cr
& {i\over\sqrt{2}}re^{-i\theta} & h_{l=1} & 0 & 0 \cr
& -{i\over\sqrt{2}}re^{i\theta} & 0 & h_{l=1} & 0 \cr
& 0 & 0 & 0 & h_{l=1} \cr}\right)
$$
where the bosonic part of the Hamiltonian is
$$
h_l = -{\gs \over 2} \left({1 \over r^5} {\partial_r} \, r^5 \partial_r
+ {1 \over r^2 \sin 4 \theta} \partial_\theta \, \sin 4 \theta \,
\partial_\theta \right) + {\gs \over 2} {l^2 \over r^2 \cos^2 2 \theta}
+ {1 \over 8 \gs} r^4 \sin^2 2 \theta \,.
$$
Note that $h_l$ is self-adjoint in the integration measure $\int r^5
\sin 4 \theta \, dr d\theta$.

There are low energy states, localized at large $r$ and small
$\theta$, which describe widely separated branes.  These states exist
because supersymmetry is unbroken, so the zero point energies
cancel.  A basis for these states is given by the wavefunctions
$$\eqalign{
\Psi_k &= {1 \over r} e^{- i k r} e^{- r^3 \theta^2 / 2 \gs} u_0 \cr
u_0 &= \left(\matrix{ & 1/\sqrt{2} \cr & -i/2 \cr
& i/2 \cr & 0 \cr }\right)\cr}
$$
Note that $\Psi_k$ becomes an exact eigenfunction as $r \rightarrow
\infty$, $\theta \rightarrow 0$, with $r^{3/2} \theta$ held
fixed.\footnote{*}{For a fixed distance $r$ between the branes,
$\Psi_k$ spreads a distance of order $\gs^{1/2}$ in the $y = r \theta$
direction.  The physical meaning of this length scale is not obvious,
as $y$ is a non-commuting coordinate.  We are grateful to Steve
Shenker for pointing this phenomenon out to us.}  It represents an
incoming $s$-wave scattering state with relative momentum $k$ and
energy ${1 \over 2} \gs k^2$.

For small $r$, the potential energy and also the off-diagonal terms in
the Hamiltonian are negligible, and the quantum mechanics problem
reduces to free motion.  This suggests that an incoming $s$-wave
scattering state could get trapped near $r=0$, and resonate there for
some time before escaping back to infinity.  This phenomenon has the
D-brane interpretation that $s$-wave scattering of two D-branes can
produce a resonant state, in which the two D-branes are bound together
by a condensate of open strings.

We can make the resonant behavior manifest with the following crude
calculation.  To find the wavefunction of the resonance, we modify the
Hamiltonian in a way which preserves its form at small $r$, but which
lifts the supersymmetric ground state.  A suitable modification is to
change the potential energy in the Hamiltonian, replacing ${1 \over 8
\gs} r^4 \sin^2 2 \theta \, u_0 u_0^\dagger$ with ${1 \over 8 \gs} r^4
\, u_0 u_0^\dagger$.  This modification prevents the scattering states
$\Psi_k$ (which are proportional to $u_0$) from escaping to
$r=\infty$, without affecting the states orthogonal to $u_0$.  It turns
the resonance into a genuine eigenstate localized near $r=0$, and we
can estimate its wavefunction.  The lowest-lying resonance has a
wavefunction
$$
\Psi_{\rm res} \approx e^{- r^3 / 6 \gs}
\left(\matrix{ & 1 \cr & A \gs^{-1/3} r \sqrt{\cos 2 \theta} \cr 
& A^* \gs^{-1/3} r \sqrt{\cos 2 \theta} \cr & 0 \cr}\right)
$$
with an energy $E_{\rm res} \approx 1.7 \, \gs^{1/3} \ms$ and $A
\approx -0.23-i0.28$ estimated by variational methods.  Note that the
resonance indeed has a size of order the eleven dimensional Planck
length $\lpii = \gs^{1/3} \ls$.

By matching this resonance wavefunction onto a superposition of
scattering wavefunctions at $r$ of order $\gs^{1/3}$, we can
obtain a crude estimate of the $s$-wave phase shift near resonance:
$$
\delta_0(k) \approx - k \lpii - {\pi \over 4} - \tan^{-1} c k \lpii
$$
where $c$ is a number of order one.  This phase shift exhibits a
resonant feature at a momentum $k$ of order $1/\lpii$.

Many of the features of the toy model should survive in the full $9+1$
dimensional problem.  We expect that a rich spectrum of resonances
should exist, with a characteristic size set by the eleven dimensional
Planck length.  But new phenomena should also arise, including the
bound states at threshold which are predicted by duality.

\bigskip
\bigskip
\centerline{\bf Note Added}
\noindent
As this work was being completed a closely related paper appeared
\DanFerSun.

\bigbreak
\bigskip
\bigskip
\centerline{\bf Acknowledgments}
\noindent
We are grateful to Tom Banks, Micha Berkooz, Mike Douglas, Greg Moore,
Valery Rubakov, Nati Seiberg and especially Steve Shenker for valuable
discussions.  This research was supported in part by DOE grant
DE-FG05-90ER40559 and by a Canadian 1967 Science Fellowship.

\listrefs
\bye